\documentclass[10pt,onecolumn]{IEEEtran}
\usepackage{color,amsfonts,amsmath,amssymb,booktabs,multirow,psfrag,bm,amsbsy,cite}
\usepackage{hyperref}
\usepackage{enumerate}
\usepackage{subfigure}
\usepackage[pdftex]{graphicx}
\usepackage{epstopdf}
  \DeclareGraphicsExtensions{.pdf,.jpeg,.png}

\begin{document}

\title{Line-of-Sight Obstruction Analysis for Vehicle-to-Vehicle Network Simulations in a Two-Lane Highway Scenario\thanks{This work was partially funded by the ELLIIT- Excellence Center at Link\"{o}ping-Lund In Information Technology and partially funded by Higher Education Commission (HEC) of Pakistan.}\thanks{T. Abbas, and F. Tufvesson are with the Department of Electrical and Information Technology, Lund University, Lund, Sweden (e-mail: taimoor.abbas@eit.lth.se; fredrik.tufvesson@eit.lth.se).}}
\author{Taimoor Abbas, and Fredrik Tufvesson}

\maketitle

\begin{abstract}
In vehicular ad-hoc networks (VANETs) the impact of vehicles as obstacles has largely been neglected in the past. Recent studies have reported that the vehicles that obstruct the line-of-sight (LOS) path may introduce $10-20$\,dB additional loss, and as a result reduce the communication range. Most of the traffic mobility models (TMMs) today do not treat other vehicles as obstacles and thus can not model the impact of LOS obstruction in VANET simulations. In this paper the LOS obstruction caused by other vehicles is studied in a highway scenario. First a car-following model is used to characterize the motion of the vehicles driving in the same direction on a two-lane highway. Vehicles are allowed to change lanes when necessary. The position of each vehicle is updated by using the car-following rules together with the lane-changing rules for the forward motion. Based on the simulated traffic a simple TMM is proposed for VANET simulations, which is capable to identify the vehicles that are in the shadow region of other vehicles. The presented traffic mobility model together with the shadow fading path loss model can take in to account the impact of LOS obstruction on the total received power in the multiple-lane highway scenarios.  
\end{abstract}

\section{Introduction}
\label{sec:Introduction}
Vehicle-to-vehicle (V2V) communication is an emerging technology that has been recognized as a key communication paradigm for safety, and infotainment applications in future intelligent transportation systems (ITS). In recent years extensive research efforts have been made to design reliable and fault tolerant vehicular ad-hoc network (VANET) communication protocols. However, the propagation channel is one of the key performance limiting factor which is not yet completely understood \cite{Javier2010}; several aspects such as the impact of antenna placement on vehicles \cite{TaimoorAWPL13} and line-of-sight obstruction by other vehicles on V2V communication has largely been neglected in the past. In \cite{TA2012arXiv,Boban11}, it is stated that a vehicle that obstructs the LOS path between the transmitter (TX) and receiver (RX) vehicle may introduce $10$\,dB additional loss in the received power, and as a result cause $3$ times reduced communication range. This additional power loss can increase up to $20$\,dB if the obstructing vehicle is tall and close to the RX vehicle \cite{Meireles10}. 

Several network simulators suitable for VANET simulations exist today, e.g., ns-2 \cite{ns-2}, OMNet++ \cite{Varga2008}, ns-3 \cite{Henderson2006}, and JiST/SWANS \cite{Barr2005}. These simulators are different from each other in terms of run-time performance and memory usage \cite{Weingartner2009}. Most of these simulators do not consider the impact of neighboring vehicles on the packet reception probabilities. To evaluate this impact in these simulators, a traffic mobility model (TMM) should be implemented having at least the ability to identify and categorize the vehicles into the following groups: 
\begin{itemize}
\item Line-of-sight (LOS) - when the TX vehicle has optical line of sight from the RX vehicle;
\item Obstructed-line-of-sight (OLOS) - when the optical LOS between the TX and RX is obstructed by another vehicle.
\end{itemize}

In the VANET simulators the role of the TMM is very vital in order to perform a realistic system simulations. Today there are a number of traffic models that can be used in the VANET simulators. Some of them are very advanced but equally complex, e.g., SUMO (Simulation of Urban Mobility) \cite{SUMO2011}, which can be implemented in any of the aforementioned VANET simulators. However, using such an advanced mobility model is not desired if the purpose of the VANET simulations is to perform a simple system analysis. Therefore, for basic packet level performance evaluations less detailed but realistic traffic flow models, e.g., the optimal velocity (OV) car-following model without or with the lane-change capabilities, \cite{Bando1995, Tang2005} respectively, can be used in the VANET simulators. 

\begin{figure*}
    \begin{center}    
    \includegraphics[width=0.85\textwidth]{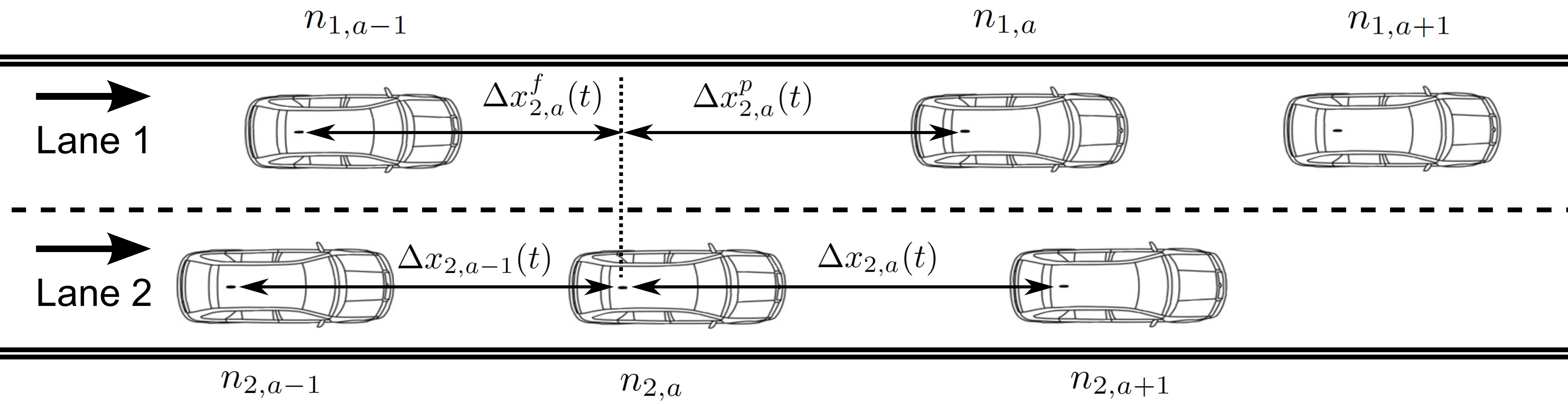}     
    \end{center}
    \caption{%
        The car-following traffic model for two-lane traffic. $\Delta x_{2,a}(t)$, $\Delta x_{2,a-1}(t)$, $\Delta x_{2,a}^p(t)$ and $\Delta x_{2,a}^f(t)$ are the headway distances from the vehicle $n_{2,a}$ to the vehicles $n_{2,a+1}$, $n_{2,a-1}$, $n_{1,a}$ and $n_{1,a-1}$, respectively. Where $n_{l,a}$ is vehicle label, $l$ is lane number and $a$ is lane specific vehicle index.
         }%
   \label{fig:Scenario}
\end{figure*}

In this paper a TMM is discussed that is capable to identify vehicles being in LOS and OLOS. The TMM is implemented in MATLAB in which the car-following model, which is used to formulate the forward motion of vehicles, is used. The car-following model is of low complexity but gives a realistic traffic flow. The interaction between the lanes is also taken into account by allowing vehicles to perform lane changes when necessary conditioned that the considered vehicles fulfill certain lane change requirements. The model is used to identify the vehicles being in LOS and OLOS from the TX at each time instant. Moreover, the instantaneous position, headway distance, state, distance traveled in each state, and number of transitions from one state to another are logged to calculate the probability of vehicles being in the LOS and OLOS states with respect to distance between them. The traffic simulations are performed based on realistic parameters and the results are compared with the measurement results collected during an independent measurement campaign (for details, see \cite{TA2012arXiv}). 

The main contribution of this paper is a TMM that is straight-forward to integrate with VANET simulators in order to study the impact of vehicles as obstruction. We do not derive the TMM itself, but we adapt models in the literature to be used for VANET simulations. As mentioned above the TMM is capable to distinguish vehicles that are in LOS and OLOS states on a two-lane highway where the traffic flow is generated by using the lane-changing rules in the car-following model. In addition to that, analytical expressions to find the packet reception probability (PRP) are also provided. The PRP can easily be estimated by utilizing the probability of being in LOS or in OLOS calculated from the TMM into the LOS/OLOS path-loss model proposed in \cite{TA2012arXiv}. Finally, the corresponding results for PRP are calculated and compared for three different V2V channel models for highway scenario; 1) the LOS only path-loss model by Karedal et~al. \cite{karedal11}, 2) the Nakagami-m based path-loss and fading model by Cheng et~al. \cite{Cheng07}, and 3) is the LOS/OLOS path-loss model in \cite{TA2012arXiv}.

The remainder of the paper is organized as follows, the TMM including the car-following model and lane change rules are discussed in Section \ref{sec:MobilityModel}. Section \ref{sec:LOSAnalysis} explains the method to distinguish between the LOS and OLOS situations. The simulation setup for the traffic mobility model, and probabilities of vehicles being in LOS and OLOS states are given in Section \ref{sec:SimulationResults}. In Section \ref{sec:PerformanceEvaluation} the analytical expressions for packet reception probabilities are analyzed for completeness, while in Section \ref{sec:summary} conclusions are given.

\section{Traffic mobility model}
\label{sec:MobilityModel}
In recent years, a number of research efforts have been made to understand and model complex traffic phenomena by using the concepts from statistical physics \cite{Chowdhury2000}. Experimental studies have also been performed to analyze traffic and lane change behaviors \cite{Xuan2012}. Among all these models, the car-following model is one of the most frequently used models to describe vehicle motion. The car-following model is capable of describing real traffic as it takes into account the velocities, headway distances, relative speeds, and the attitude of the drivers to model the traffic flow. The optimal velocity (OV) car following model, first introduced by Bando et~al. \cite{Bando1995}, was extended for two-lanes in \cite{Tang2005}. Tang et~al. \cite{Tang2007} further extended the model to incorporate the effect of potential lane changing and analyzed the traffic flow stability. The car following model for two-lane traffic flow is discussed underneath, in which the lane changing is also allowed. The model is modified such that the probabilities of vehicles being in LOS and OLOS situations can be obtained using simple geometric manipulations that can further be integrated into the VANET simulators. 
\subsection{The car-following model}
\label{subsec:LaneChangeRules}
Consider a highway with two lane traffic in each direction of travel and assume that the vehicles in each lane move along a straight line. Let $l = \{1,2\}$ be the lane index for the outer (fast) and inner (slow) lane, respectively. Vehicles in lane $l$ are labeled as $(...,n_{l,a-1},n_{l,a},n_{l,a+1},...)$, where $a$ is a lane specific vehicle index, their instantaneous positions are $(...,x_{l,a-1}(t),x_{l,a}(t),x_{l,a+1}(t),...)$ and the headway between any two vehicles moving in the same lane are labeled as $(...,\Delta x_{l,a-1}(t),\Delta x_{l,a}(t),\Delta x_{l,a+1}(t),...)$ at time instant $t$, as described in Fig.~\ref{fig:Scenario}. At each time instant $t$ each of the two lanes will be classified as subject-lane or target-lane with respect to each subject vehicle. A subject-lane is the lane where the vehicle $n_{l,a}$ drives, and target-lane is the lane on which the vehicle $n_{l,a}$ intend to drive after the possible lane change.

\begin{figure*}
    \begin{center}    
    \includegraphics[width=0.9\textwidth]{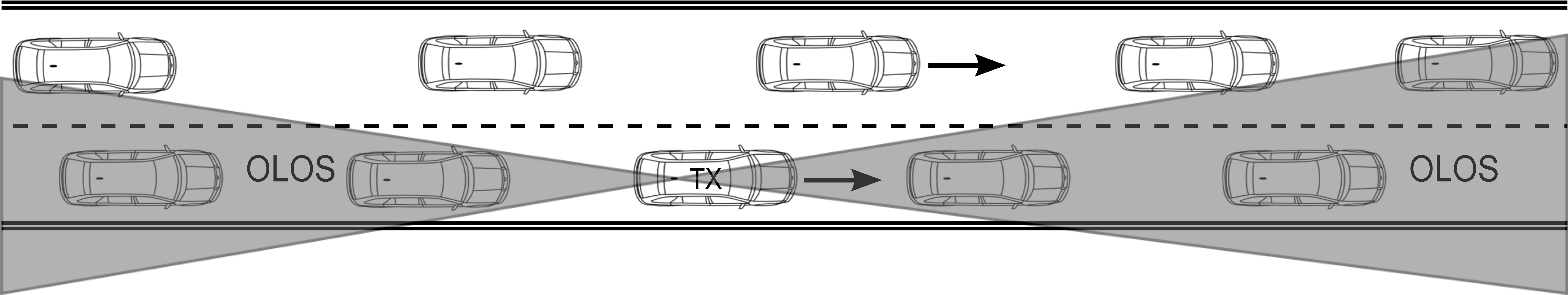}     
    \end{center}
    \caption{%
        Identification of vehicles being in LOS and in OLOS of the TX vehicle; vehicles in the shaded-area are considered to be in OLOS where as all other vehicles have LOS from the TX.
    }%
   \label{fig:LOS_OLOS_Identification}
\end{figure*}

A microscopic simulation model, the car-following model, is used to describe the movement of vehicles on the same lane. It explains a one-by-one following process of vehicles and incarnate human behaviors which in turn reflects realistic traffic conditions. It has been shown that the car-following model is a better way to model traffic-flow compared to the other common traffic-flow models \cite{Yanlin2001}. Tang et al. \cite{Tang2005, Tang2007} developed a car-following model for two-lane traffic-flow in the forward direction, expressed as follows: 

\begin{eqnarray}
\label{eq:car-following}
\frac{d^2 x_{l,a}(t)}{dt^2} &=& \alpha_l \left(V_l \left(\Delta x_{l,a}(t),\Delta x_{l,a}^p(t)\right)-\frac{dx_{l,a}(t)}{dt}\right) \nonumber\\
&& +\kappa_l\Delta v_{l,a}(t),
\end{eqnarray}
where $\Delta v_{l,a}(t)$ is the relative velocity between two vehicles $n_{l,a}$ and $n_{l,a+1}$, $\Delta x_{l,a}^p(t)$ is the distance between the vehicle $n_{l,a}$ and the preceding vehicle in the target lane, $\alpha_l$ is the driver's sensitivity coefficient, and $k_l = \frac{\lambda_l}{\tau_l}$ is the sensitivity coefficient due to difference in velocity, in the lane $l$ at time instant $t$, respectively. The delay $\tau_l$ is the time delay in which a vehicle attains its optimal velocity and $\lambda_l \in (0,1)$ is the sensitivity factor for the relative velocities which is independent of time, position and velocity. However, it is assumed that the driving condition is better in the outer (fast) lane $1$ compared to the inner (slow) lane $2$, and thus $\lambda_1>\lambda_2$. 

The continuous model in (\ref{eq:car-following}) can be discretized using forward difference to find the position of vehicle $n_l$ at any time $t+2\tau_l$ \cite{Tang2007} as given below,

\begin{eqnarray}
x_{l,a} (t+2\tau_l) &=& x_{l,a}(t+\tau_l)+\tau_lV_l\left(\Delta \tilde{x}_{l,a}(t)\right)+ \nonumber\\
&& \lambda_l\tau_l\left(x_{l,a}(t+\tau_l)-x_{l,a}(t)\right).
\label{eq:position}
\end{eqnarray}
The above equation can also be written in terms of headways as,

\begin{eqnarray}
\Delta x_{l,a} (t+2\tau_l) &=& \Delta x_{l,a}(t+\tau_l) \nonumber\\
&& +\tau_lV_l\left(\Delta \tilde{x}_{l,a+1}(t)\right)-V_l\left(\Delta \tilde{x}_{l,a}(t)\right) \nonumber\\
&& +\lambda_l\tau_l\left(x_{l,a+1}(t+\tau_l)-x_{l,a+1}(t)\right) \nonumber\\
&& -\left(x_{l,a}(t+\tau_l)-x_{l,a}(t)\right), 
\label{eq:headways}
\end{eqnarray}
where $V_l\left(\Delta \tilde{x}_{l,a}(t)\right)$ is the headway induced optimal velocity function (OVF). The OVF is given as follows,
\begin{eqnarray}
V_l\left(\Delta \tilde{x}_{l,a}(t)\right) = V_l\left(\Delta x_{l,a}(t),\Delta x_{l,a}^p(t)\right)  \nonumber\\
= \frac{1}{2} v_{l,max}\left(\tanh(\tilde{x}_{l,a}(t)-d_l^p)+\tanh(d_l^p)\right),
\label{eq:OVF}
\end{eqnarray}
where $d_l^p$ and $v_{l,max}$ are the minimum safety distance from the preceding vehicle in the target lane and the maximum velocity in the lane $l$, respectively. Finally the weighted headways $\tilde{x}_{l,a}(t)$ are defined as,
\begin{equation}
\tilde{x}_{l,a}(t) = \beta_1 \Delta x_{l,a} + \beta_2 \Delta x_{l,a}^p(t),
\label{eq:weightedheadways}
\end{equation}
where $\beta_1$ and $\beta_2$ are the weights for the headways from the preceding vehicles in same lane and the target lane, respectively, and $\beta_1>\beta_2$ given that $\beta_1+\beta_2=1$. The car following model explained above is used to formulate the forward motion of vehicles. 

The forward difference equations, (\ref{eq:position}) and (\ref{eq:headways}), used to find the positions and headways of the vehicles, respectively, do take the driver sensitivity coefficients and sensitivity factor for the relative velocity into account. However, many other factors (e.g., weather condition, road bumps, and driver mood) can also influence the traffic flow. Moreover, the vehicles are assumed to be moving along a straight line, which means no variations along the vertical axis and is not the case in reality. To summarize, we can say that the generated traffic flow is realistic but due to simplifications it is noise free in the sense that the vehicles follow the center point of the lanes. Hence, it is important to introduce some randomness to make the result of the TMM more realistic, which is done in section \ref{sec:PerformanceEvaluation} when the TMM is integrated with the LOS/OLOS path-loss model.

\subsection{Lane change rules}
\label{subsec:LaneChangeRules}
To characterize realistic traffic in a multi-lane highway scenario it is important to consider interaction between lanes and the lane change activities as it affects stability of the traffic flow. In \cite{Tang2007} it is concluded that if lane changes are not allowed then the system has a stable flow, but when the vehicles are allowed to change lanes then the system flow can become metastable or unstable depending upon the frequency of lane change activities.

In our simulator each vehicle is allowed to perform lane changes when necessary, conditioned that the vehicle fulfills all lane change requirements. During a lane change event both the lanes are categorized either as the subject-lane or the target-lane. Whenever a vehicle changes lane from the subject to target lane it becomes a vehicle in the target lane, and thus the position, number and identity of each vehicle in both lanes are updated accordingly. It is assumed that the lane change process is instantaneous, so when a vehicle changes lane its longitudinal location remains the same as it was prior to the lane change.   

In \cite{Kurata2003,Tang2007,Sheng2009} several lane changing rules are defined that can either be used independently or all together to model the lane change behavior. The lane-changing rule based on the incentive and safety criterion defined in \cite{Tang2007} states that the vehicle is allowed to change lane only if it fulfills the following three criteria,

\begin{itemize}
\item The distance of the vehicle $n_{l,a}$ from the preceding vehicle $n_{l,a+1}$ should be smaller than twice the safety distance $d_l^p$, i.e.,
\begin{equation}
\Delta x_{l,a}(t)<2d_l^p.
\label{eq:rule1}
\end{equation}
\item The distance of the vehicle $n_{l,a}$ from the preceding vehicle in the target lane should be greater than the distance of the vehicle $n_{l,a}$ from the preceding vehicle $n_{l,a+1}$ in the same lane, i.e.,
\begin{equation}
\Delta x_{l,a}^p(t)>\Delta x_{l,a}(t).
\label{eq:rule2}
\end{equation}
\item Finally the distance $\Delta x_{l,a}$ of the vehicle $n_{l,a}$ from the vehicle in the target lane following this vehicle $n_{l,a}$ should be greater than the corresponding safety distance of the following vehicles $d_l^f$, i.e., 
\begin{equation}
\Delta x_{l,a}^f(t)>d_l^f
\label{eq:rule3}
\end{equation}
\end{itemize}

In \cite{Filzek2001} it is stated that $0.9$\,s is the minimum legal time-gap during following, which gives the safety distance relative to the velocity of the vehicle. Their measurement results show that the time-gap during following is not fixed but it is relative to the speed of the vehicle and traffic density. Thus, we can say that the safety distance $d_l^p$ and the corresponding safety distance of the following vehicles $d_l^f$ are random parameters which depend on the velocity of the subject vehicle given a minimum time-gap. In general the so-called two-second rule is a rule of thumb to determine the correct following distance, i.e., a driver should ideally keep at least two seconds of time-gap from any vehicle that is in front of the subject vehicle.
\section{Line-of-sight obstruction analysis}
\label{sec:LOSAnalysis}

As mentioned above, to date most of the VANET simulators do not consider the impact of line-of-sight obstruction, caused by neighboring vehicles, on the packet reception probabilities. To evaluate this impact in the simulator the TMM is required to identify and label each vehicle as in LOS or in OLOS situation with respect to TX and RX at each instant $t$. The identification of vehicles being in LOS or in OLOS states becomes fairly simple as the TMM discussed earlier provides the instantaneous position of each vehicle on the road. Thus, the position information of each vehicle together with some geometric manipulations give the state information of each vehicle being in LOS or in OLOS state as follows,

\begin{itemize}
\item Model each vehicle as a screen or a strip with the assumption that each vehicle has the same size.
\item Assumed that the \emph{intended} communication range is a circle of a certain radius, i.e., $R_c$. At each instant $t$ the vehicles that are in this circle are considered only.
\item Vehicles in each lane are assumed to be moving along a straight line. Thus only two vehicles in the same, one at the front and one in the back of the TX, will be in the LOS. The rest of the vehicles in the same lane are considered to be in the OLOS state.
\item Draw straight lines starting from the antenna position of the TX vehicle touching the edges of the vehicles in the front and back to the edges of road (see Fig.~\ref{fig:LOS_OLOS_Identification}). All vehicles that are bounded by these lines are shadowed by other vehicles thus in the OLOS state. 
\item Vehicles that are not bounded by these lines are analyzed individually to see if they are in LOS or in OLOS from the TX. 
\item The identification process is repeated for each vehicle and at each time instant $t$ to find out whether the vehicles are in LOS and in OLOS states with respect to every other vehicle. The state information of each vehicle can then be used either for analytical performance evaluations or for packet level VANET simulations. 
\end{itemize}
\section{Simulations and Results}
\label{sec:SimulationResults}

The TMM derived above is implemented in Matlab and simulations are carried out in order to analyze the movement of vehicles over time, their lane changing behavior and the intensities by which the vehicles change states from LOS-to-OLOS and from OLOS-to-LOS states, respectively. The simulations are performed on a two-lane  $14.4$\,km long circular highway. The circular highway refers to the fact that any vehicle that departs from one end of the highway, i.e., beyond $14.4$\,km, enters from the other end so that the traffic can flow for infinite amount of time. The simulation parameters are chosen as follows.

\begin{figure}
     \begin{center}     
     \includegraphics[width=.5\textwidth]{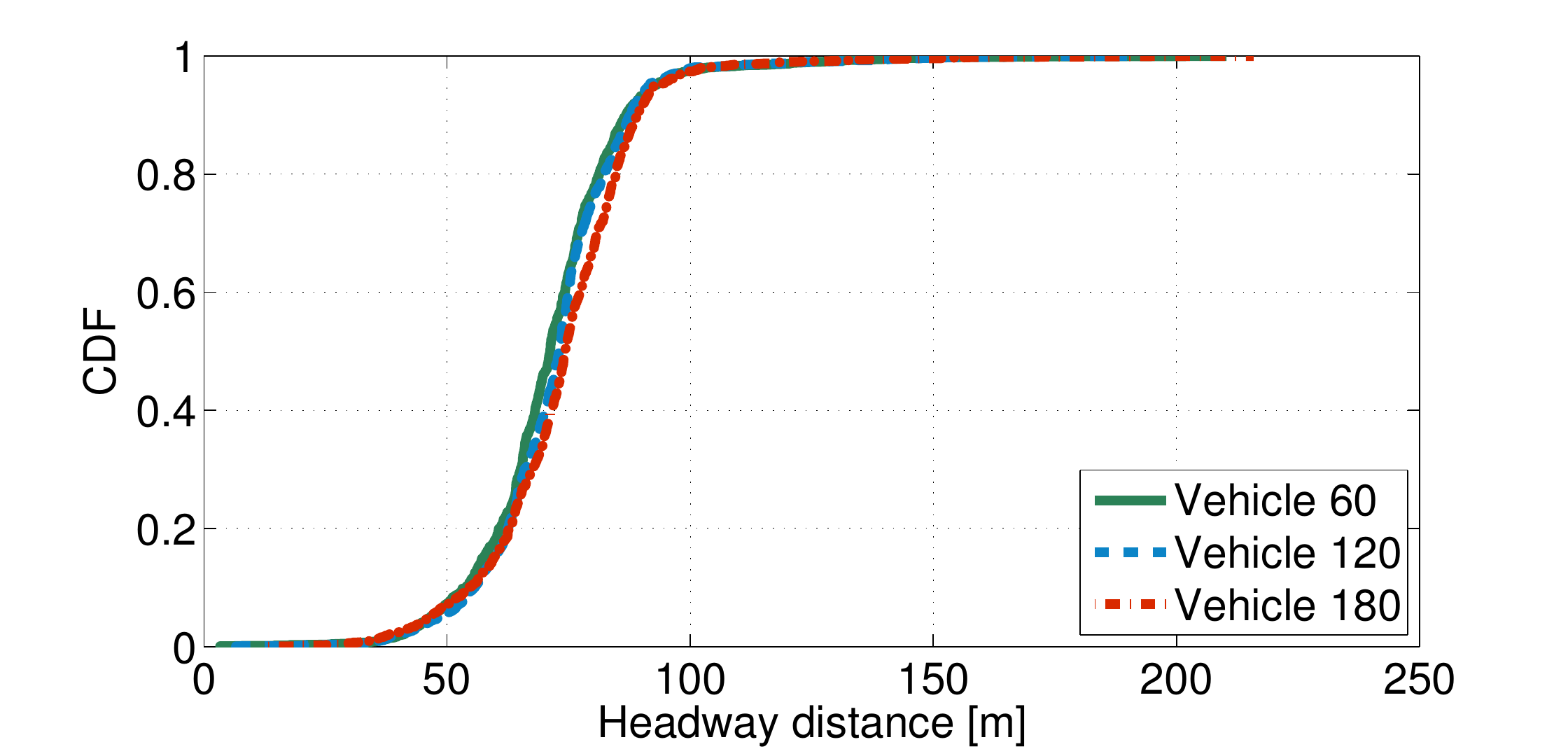}
     \end{center}
     \caption{CDFs of the headway distances of vehicles at every second for total simulation time $T=120$\,min.}%
   \label{fig:Headways}
\end{figure}

For the simulations, the initial positions $x_{l,a}(0)$ and the headways $\Delta x_{l,a}(0)$ of all the vehicles $n_{l,a}$ in lane $l$ for ($a = 1,2,...,N_l$) are determined by the rules given in \cite{Tang2007} for both the lanes, $l=\{1,2\}$. Initially it is assumed that the vehicles are distributed uniformly along each lane with the realistic flow rate given in the Highway Capacity Manual \cite{Capacity1998}, i.e., $1300$\,vehicles/hour/lane and $1600$\,vehicles/hour/lane at an average speed of $30.5$\,m/s ($110$\,km/h) and $22.5$\,m/s ($80$\,km/h) in the outer-lane $l=1$ and inner-lane $l=2$, which implies $1$\,vehicle per $3$\,s and $1$\,vehicle per $2.5$\,s, respectively. 

The initial values of $\Delta x_{l,a}^p(0)$ and $\Delta x_{l,a}^f(0)$ are determined from the initial positions $x_{l,a}(0)$ of the vehicles. The position and headways at each instant are updated by (\ref{eq:position}) and (\ref{eq:headways}).

Let $N_1=160$ and $N_2=200$ be the initial number of vehicles in each lane, $v_1=27.7$\,m/s ($100$\,km/h) and $v_2=19.44$\,m/s ($70$\,km/h) be the average velocity, and $v_{1,max}=30.5$\,m/s ($110$\,km/h) and $v_{2,max}=22.2$\,m/s ($80$\,km/h) be the maximum speed in the outer and inner lanes, respectively. The other parameters such as the delay time, sensitivity factors, and initial safety distances are $\tau_1=\tau_2=0.5$\,s, $\lambda_1=0.3$ and $\lambda_2=0.2$, and $d_1^f=d_1^p=40.5$ and $d_2^f=d_2^p=36$\,m, respectively. 

\begin{figure}
    \begin{center}    
    \includegraphics[width=0.5\textwidth]{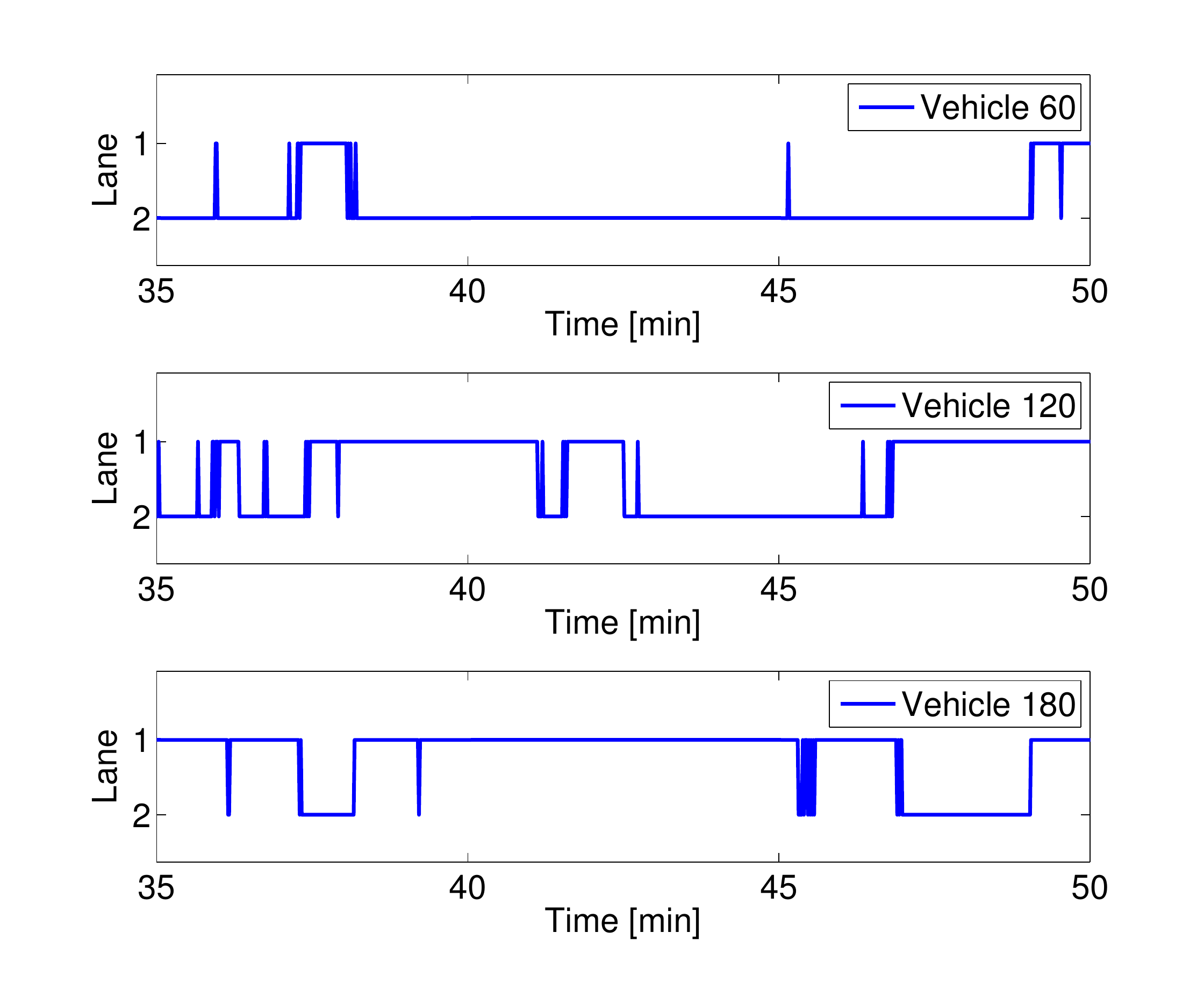}     
    \end{center}
    \caption{%
        Three vehicles numbered $60, 120$ and $180$ changing lanes from lane $1$ to lane $2$ or vice-versa between a time window of $35$\,min to $50$\,min.
    }%
   \label{fig:LaneChanges}
\end{figure}

Practically, the driver's sensitivity $\alpha_1$ is larger than $\alpha_2$ because the driver's response in the outer (fast) lane is more sensitive than in the inner (slow) lane. Here we assume that $\alpha_1=\alpha_2$ because for the simulations it is easy to compute headways at fixed intervals and it is anticipated in \cite{Tang2007} that the effect of $\alpha_l$ is small and does not change final results. 

We let the simulations run for $10800$ simulation time steps or seconds that correspond to $3$\,hours of simulated time. The data obtained from the first $3600$\,s of simulation is not considered for analysis to ensure that steady-state conditions are obtained. Hence, the time $0$\,s in the final results corresponds to the time $3600$\,s of the simulation. 

Once the traffic flow is stable the positions and headways of all the vehicles are logged for each time instant, for further analysis, with respect to the vehicles' identity. The vehicles are allowed to change lane so whenever a vehicle changes lane it exits from subject lane and becomes part of the target lane. Thus for every lane change event at each time instant $t$ the position, and headway distances of each vehicle in both lanes, the subject and target lanes, should be updated accordingly. 

The headways for three vehicles numbered $60, 120$ and $180$ are shown as cumulative distribution function (CDF) in Fig.~\ref{fig:Headways}. It can be seen that there is a huge variation in the headway distances and they may vary between $20$\,m up to $600$\,m. 

Further, to record the lane change activities, the total number of lane changes, the position and time at which lane change occurred were logged over the simulation time for each vehicle. A sample result is shown in Fig.~\ref{fig:LaneChanges} where the lane change activities of the three vehicles numbered $60, 120$ and $180$ are shown over $15$\,min of time window. It can be seen that the lane change behavior for each vehicle is different at different times. The amount of time a vehicle stays in each lane depends very much on the driving conditions in that lane during that particular time window.

\begin{figure}
    \begin{center}    
    \includegraphics[width=0.5\textwidth]{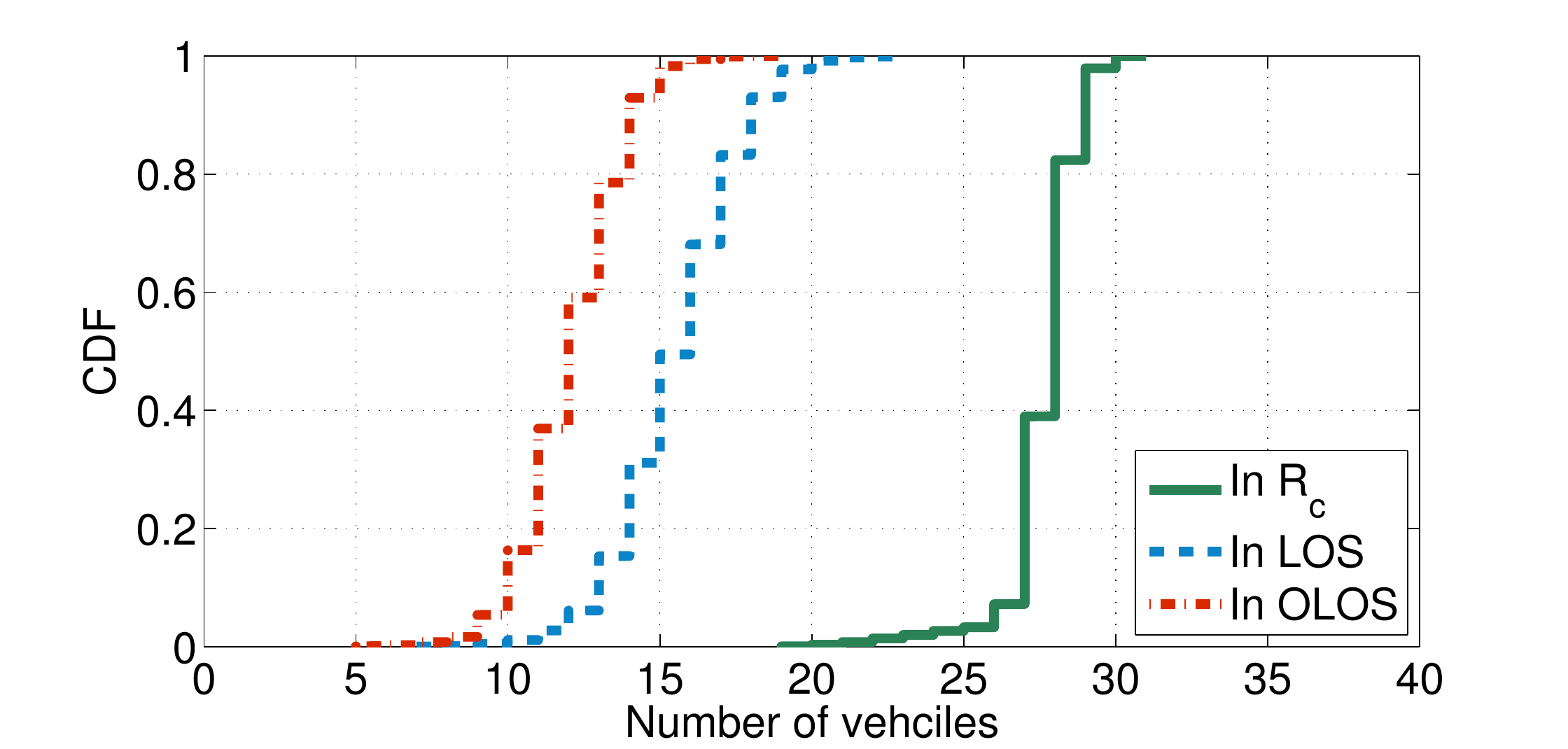}     
    \end{center}
    \caption{%
        CDFs of the total number of vehicles in $R_c$ and, the number of vehicles in LOS and OLOS state at each time instant for total simulation time $T=120$\,min. 
    }%
   \label{fig:States}
\end{figure}
\begin{figure}
     \begin{center}
        \subfigure[]{%
            \label{fig:Intervals_LOS_OLOS}
        		\includegraphics[width=.5\textwidth]{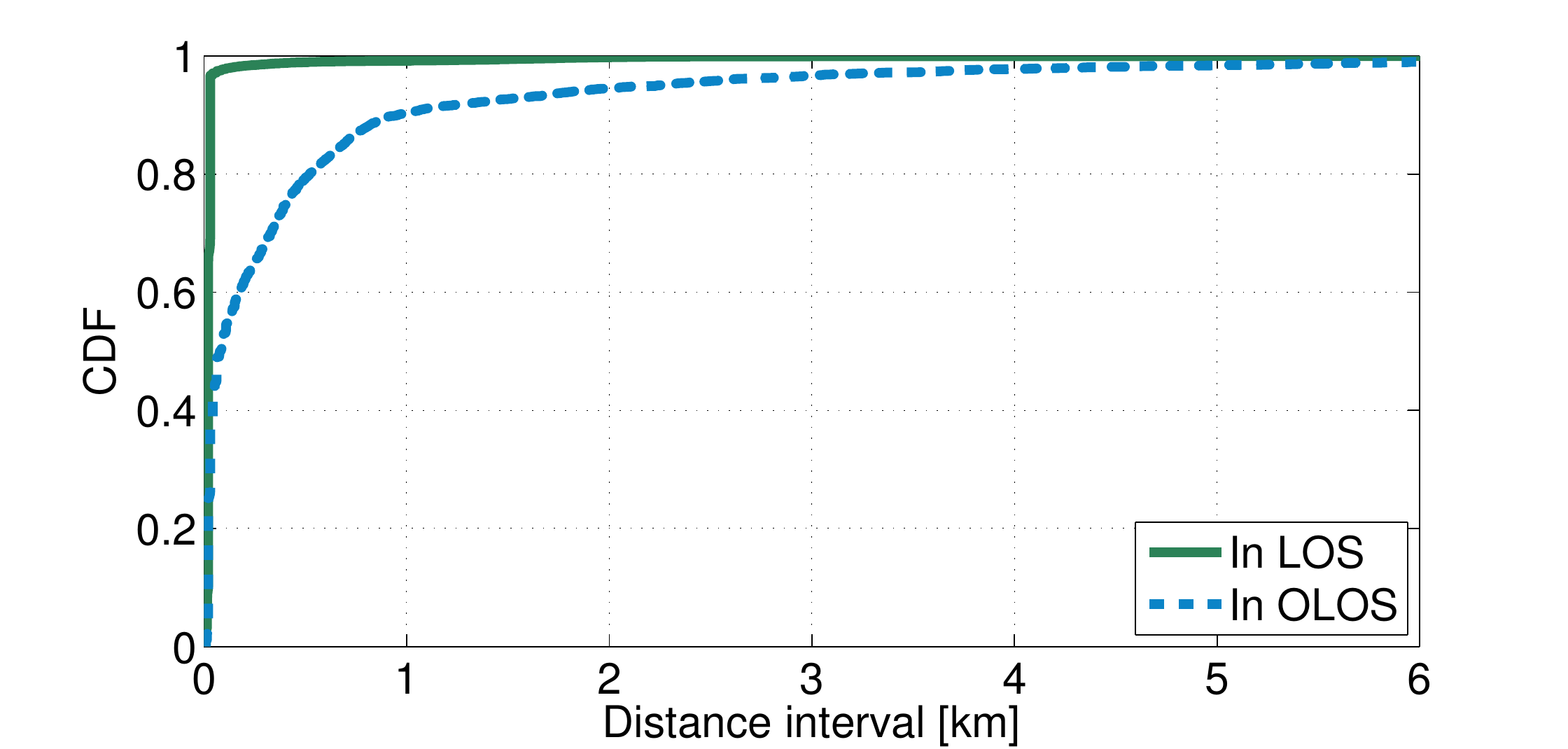}%
  			}
  			\\
        \subfigure[]{%
            \label{fig:Distance_traveled_LOS_OLOS}
            \includegraphics[width=.5\textwidth]{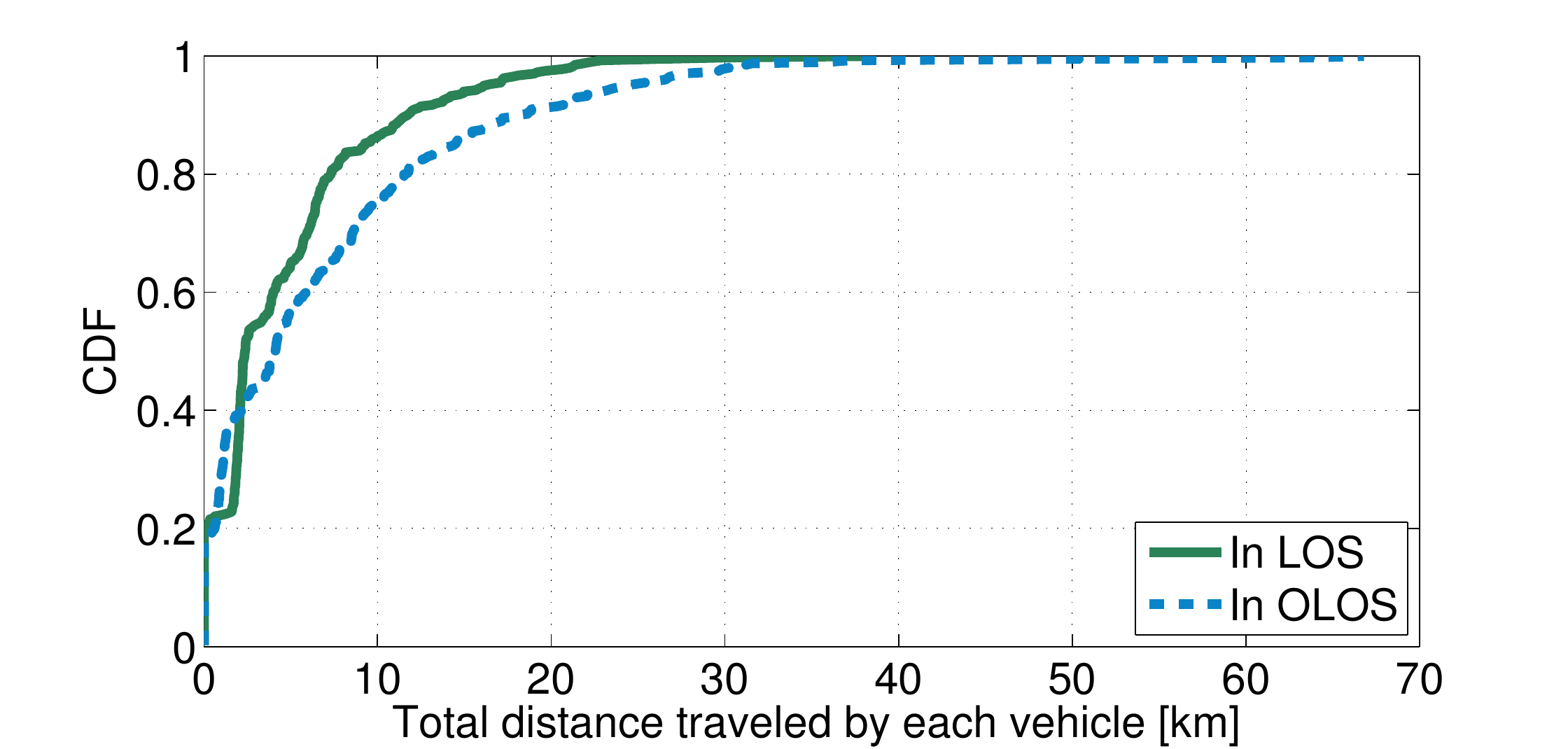}
        }
          			        
    \end{center}
    \caption{CDFs of; (a) the LOS and OLOS intervals for all vehicles, and (b) the total distance traveled in the LOS and OLOS by all vehicles.}%
   \label{fig:Intervals_Distance_LOS_OLOS}
\end{figure}

As the main focus of this work is to identify the vehicles which are in OLOS from each other so that this information can be used for VANET simulations using the shadow fading path loss model given in \cite{TA2012arXiv}. In order to analyze the LOS and OLOS situation and to find the intensities by which vehicles go from one state to another the following assumptions are made:

A vehicle numbered $20$ is assumed to be the TX vehicle which is broadcasting the information with in the intended communication range $R_c$ where $R_c$ is a circle of radius $500$\,m with TX at its center. At each instant $t$ the vehicles which lie in the $R_c$ of the TX vehicle are identified and then categorized as vehicles being in LOS or in OLOS from the TX vehicle using the rules defined in section III. Any other vehicle that is outside this intended communication range $R_c$ is treated as a vehicle out-of-range (OoR) from the TX. The states of vehicles being in LOS, OLOS and OoR w.r.t. their identities are saved for each time instant. The CDF of the total number of vehicles in $R_c$ and, the number of vehicles in LOS and OLOS state at each time instant are shown in Fig.~\ref{fig:States}, respectively. The OoR state is not interesting thus it is not discussed further.

\begin{figure}
\begin{center}
\includegraphics[width=.5\textwidth]{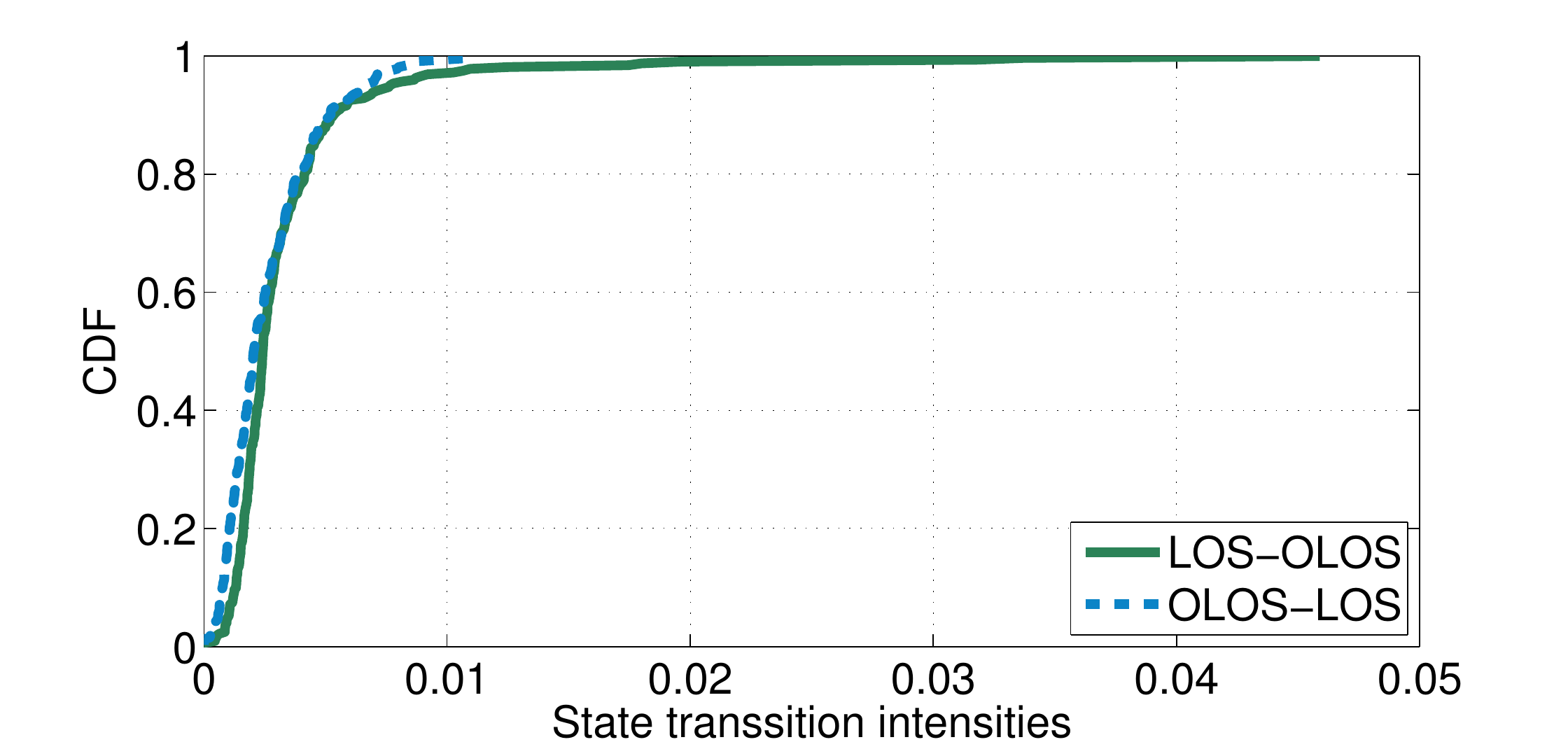}%
\end{center}
\caption{CDFs of the state transition intensities $P$ and $p$ from LOS-OLOS and OLOS-LOS, for each vehicle, respectively.}%
\label{fig:Transitions_Intensities_LOS_OLOS}
\end{figure}

\begin{figure}
\begin{center}
\includegraphics[width=.5\textwidth]{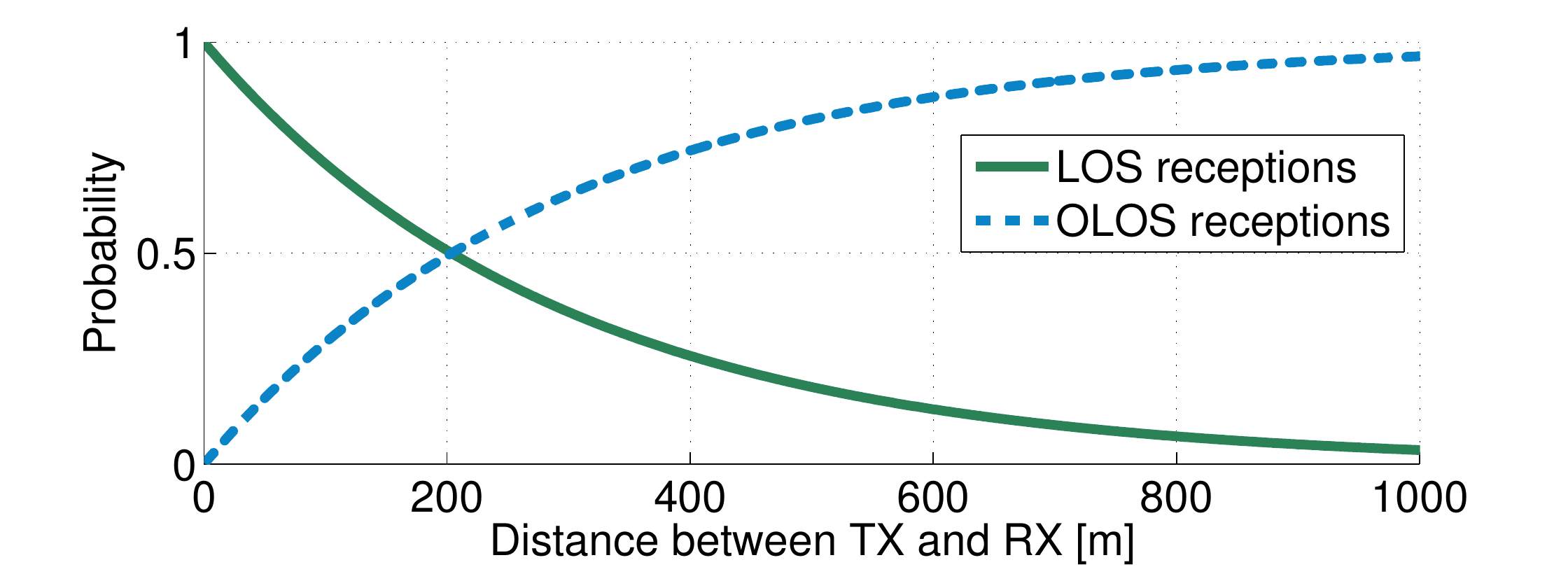}%
\end{center}
\caption{The probability of LOS and OLOS with respect to distance, it can be seen that the probability of being in LOS decreases as the distance increases.}
\label{fig:Probability_LOS_OLOS}
\end{figure}


Each time a vehicle is in LOS, or in OLOS, it remains in that state for a certain amount of time and travels a distance, $d_{l,a}^{LOS}(k)$ or $d_{n_l,a}^{OLOS}(k)$, where $k \in \{1,K\}$ is the index of that specific interval. The length of these intervals may vary over time as well as for each vehicle. So we log the count of these intervals and their corresponding distances $d_{l,a}^{LOS}(k)$ and $d_{n_l,a}^{OLOS}(k)$ for every vehicle over the whole simulation time. The CDFs of LOS and OLOS distance intervals for all vehicles are shown in Fig.~\ref{fig:Intervals_LOS_OLOS}. We log the total distance traveled by each vehicle, $D_{l,a}$, during the simulation time and see how much of that distance is traveled in the LOS and OLOS state, $D_{l,a}^{LOS}$ and $D_{l,a}^{OLOS}$, by the vehicle $n_{l,a}$. The CDFs of total distance traveled in the LOS and OLOS by all vehicles are shown in Fig.~\ref{fig:Distance_traveled_LOS_OLOS}.

The number of state transitions, $N_{l,a}^{LOS-OLOS}$ and $N_{l,a}^{OLOS-LOS}$, from LOS-OLOS and OLOS-LOS states are counted for each vehicle. Thus the state transition intensities $P$ and $p$ from LOS-OLOS and OLOS-LOS for each vehicle can be calculate as, 
\begin{equation}
P= \frac{N_{l,a}^{LOS-OLOS}}{D_{l,a}^{LOS}},
\label{eq:P}
\end{equation} 
\begin{equation}
p= \frac{N_{l,a}^{OLOS-LOS}}{D_{l,a}^{OLOS}}.
\label{eq:p}
\end{equation} 

The CDF of the state transition intensities $P$ and $p$ for a given set of parameters are shown in Fig.~\ref{fig:Transitions_Intensities_LOS_OLOS}. The variations in the transition intensities are due to the fact that each vehicle has different moving and lane-changing pattern. The mean intensities $\mu_{P}$ and $\mu_{p}$ are calculated to be $0.0034$\,$m^{-1}$ and $0.0026$\,$m^{-1}$, respectively. For comparison, sample state transition intensities are also calculated from the measurement data collected during a V2V measurement campaign conducted in the city of Lund and Malm\"{o}, Sweden, to analyze the shadow fading effects. The measurement data was separated for LOS and OLOS conditions (explained briefly in \cite{TA2012arXiv}). The separated data contains information about the number of state transitions between LOS and OLOS states, and the distance traveled in each state. With this information the state transition intensities are calculated using (\ref{eq:P}) and (\ref{eq:p}), i.e., $P_{measured}=0.0035$\,$m^{-1}$ and $p_{measured}=0.0020$\,$m^{-1}$, which are close to the mean values of the simulated intensities. The probability of vehicles being in LOS and in OLOS with respect to the distance can also be calculated from the simulation, as shown in Fig.~\ref{fig:Probability_LOS_OLOS}.

\section{Analytical Performance Evaluation}
\label{sec:PerformanceEvaluation}
In order to evaluate the impact of vehicle as an obstruction on V2V networks the proposed TMM together with the LOS/OLOS path-loss model given in \cite{TA2012arXiv} can be used in any VANET simulator. The LOS/OLOS path-loss model provides the deterministic and stochastic parameters of a dual slope distance dependent path-loss for both the LOS and OLOS situations. The stochastic part of the LOS/OLOS path-loss model comes from the large-scale fading, which is assumed to be Gaussian distributed. The packet reception probability (PRP) can be obtained by analytical expressions for all vehicles either in LOS or in OLOS states. Large-scale fading, or shadow fading, may refer to the signal variations that may not only be associated to blocked LOS but due to the blocking of many other significant reflected propagation paths. Therefore, it is associated to both the LOS and the OLOS state. The large-scale fading is a random process and it varies over time due to varying locations when the TX/RX vehicles are moving. The proposed TMM is assumed to be noise free, therefore the required noise due to randomness in driving behavior can be taken into account by large-scale fading process, which has a standard deviation $\sigma$, that introduces variation in the received power due to variation in the position of each vehicle at each instant.

To study the performance differences in the PRP with and without considering vehicles as obstacles the LOS/OLOS model is compared with two other aforementioned path-loss models; 1) the LOS only single slope path-loss model by Karedal et~al. \cite{karedal11}, 2) the Nakagami-m based path-loss and fading model by Cheng et~al. \cite{Cheng07} in which the data from LOS and blocked LOS cases is lumped together for modeling purpose.

\begin{figure}
\begin{center}
\includegraphics[width=.5\textwidth]{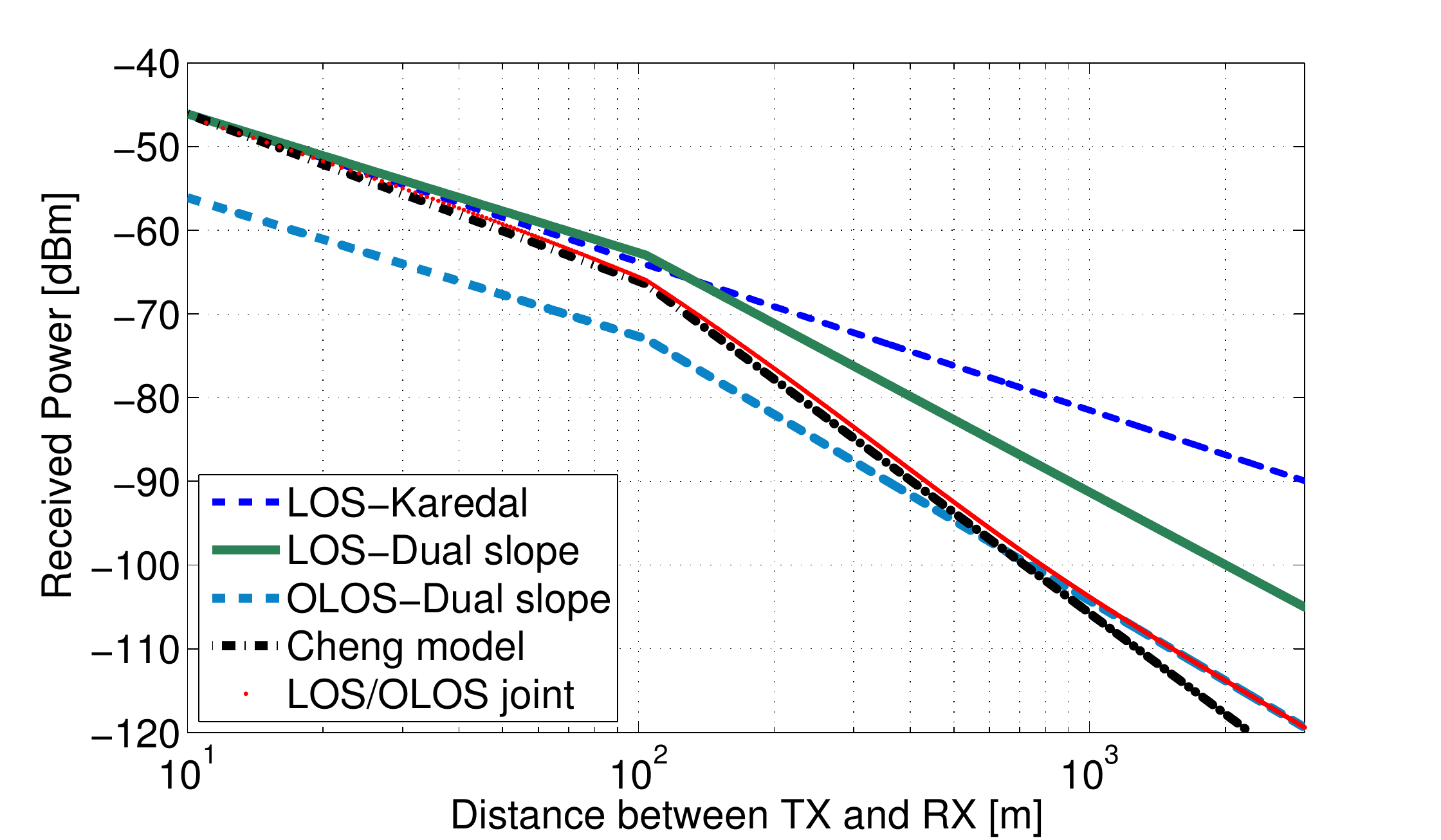}%
\end{center}
\caption{Received power as a function of distance. Breakpoint distance of $d_b=104$\,m is used for the LOS-Dual slope, OLOS-Dual slope, Cheng and joint LOS/OLOS models.}
\label{fig:Pathloss}
\end{figure}

To find an analytical expression for packet reception probability, it is assumed that each vehicle is a point source and vehicles are distributed along a straight line on both lanes of the highway and the probability of LOS and OLOS is known. The parameters of Karedal's LOS model, Cheng's Nakagami based model, and LOS/OLOS model are taken from \cite{karedal11}, \cite{Cheng07}, and \cite{TA2012arXiv}, respectively. Then the received power $Pw_{RX}$ for LOS-Karedal, LOS-Dual slope, OLOS-Dual slope, Cheng model, and joint LOS/OLOS (LOS/OLOS model together with probability of LOS and OLOS) cases can individually be calculated as follows,

\begin{equation}
Pw_{RX}(d)=Pw_{TX}-PL(d),
\label{eq:RxPower}
\end{equation}
where $PL(d)$ is a distance dependent mean power loss, given as,

\begin{equation}
    PL(d)= 
\begin{cases}
    PL_0+10n_1\log_{10}\left(\frac{d}{d_0}\right)+X_{\sigma},& \text{if } d_0\le d\le d_b\\
    PL_0+10n_1\log_{10}\left(\frac{d_b}{d_0}\right)+ & \text{if } d > d_b\\
    		10n_2\log_{10}\left(\frac{d}{d_b}\right)+X_{\sigma},
\end{cases}
\label{eq:dual-slope-PowerLaw}
\end{equation}
where $X_\sigma$ describes the large scale fading as zero mean Gaussian distributed random variable with standard deviation $\sigma$, $PL_0$ is the received power level at a reference distance $d_0=10$\,m and, $n_1$ and $n_2$ are the path-loss exponents, respectively. The value of $PL_0, n_1, n_2$ and $\sigma$ for each of the above mentioned models are different and are obtained from the models given in \cite{karedal11,Cheng07,TA2012arXiv}. The received power for all five cases is shown in Fig.~\ref{fig:Pathloss} for a transmitted power $Pw_{TX}=20$\,dBm. For the dual-slope LOS/OLOS model and Cheng's model the break point distance is provided, i.e., $d_b=104$\,m, however for Karedal's single slope LOS model $d_b$ is not required and thus it can assumed to be infinity.

From the above equations it is obvious that the received power is a Normally distributed with a distance dependent mean $\mu(d)=Pw_{TX}-PL(d)$ and standard deviation $\sigma$. The Gaussian probability density function is closely related to Q-function \cite{Proakis2006}, therefore, for a given distance $d$ the probability of received power being greater than $\alpha,$ $P\{Pw_{RX}(d) > \alpha\}$, is calculated analytically as follows,

\begin{equation}
P\{Pw_{RX}(d) > \alpha\}=1-Q\left(\frac{\mu(d)-\alpha}{\sigma}\right),
\label{eq:largerthan_alpha}
\end{equation} 
where $\alpha$ is carrier sense threshold (CSTH). The parameters for each of these models can be used individually to find the probabilities $P^{Karedal}\{Pw_{RX}(d) > \alpha\}$, $P^{Cheng}\{Pw_{RX}(d) > \alpha\}$, $P^{LOS}\{Pw_{RX}(d) > \alpha\}$ and $P^{OLOS}\{Pw_{RX}(d) > \alpha\}$, respectively. 

\begin{figure}
\begin{center}
\includegraphics[width=.5\textwidth]{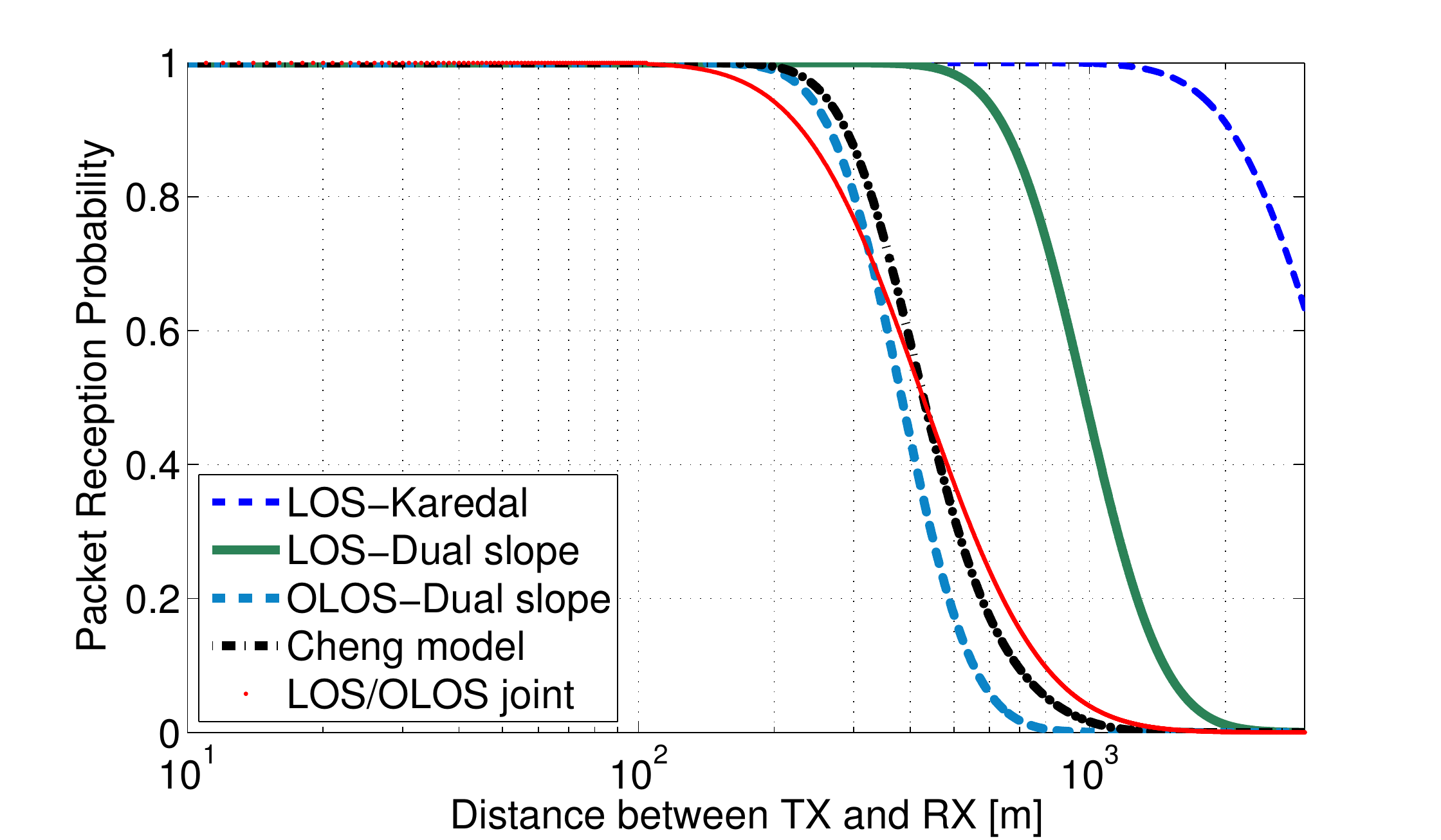}%
\end{center}
\caption{The probability of successful packet reception for a CSTH of $-91$\,dBm.}
\label{fig:PRP}
\end{figure}

The probability of successful packet reception is shown in Fig.~\ref{fig:PRP}, where CSTH$=-91$\,dBm is assumed \cite{Katrin2011}. However the joint LOS/OLOS PRP is calculated by multiplying the probability of LOS and OLOS to the individual PRP, $P^{LOS}\{Pw_{RX}(d) > \alpha\}$ and $P^{OLOS}\{Pw_{RX}(d) > \alpha\}$, of LOS and OLOS as follows,
 
\begin{eqnarray}
PRP^{LOS/OLOS} = Pr^{LOS}\times P^{LOS}\{Pw_{RX}(d) > \alpha\} + \nonumber\\
Pr^{OLOS}\times P^{OLOS}\{Pw_{RX}(d) > \alpha\}.
\label{eq:largerthan_alpha}
\end{eqnarray} 

From the Fig.~\ref{fig:Pathloss}~and Fig.~\ref{fig:PRP}, it can obviously be seen that the LOS and OLOS situations are fundamentally different. Comparing the PRP curves from the Karedal, Cheng and LOS/OLOS model, it can be observed that for the given vehicular traffic density the probabilities of LOS and OLOS vary which in turn affect the performance of the joint LOS/OLOS PRP. However, Karedal's path-loss model and Cheng's model do not take probabilities of LOS and OLOS into account and thus can not capture the effects of traffic density on the PRP. All models perform similar up to $d=100$\,m approximately. However at the larger distances, where the probability of LOS obstruction increases, the behavior of these models differ.

\section{Summary and Conclusions}
\label{sec:summary}
In this paper the effect of line-of-sight (LOS) obstruction is analyzed for vehicle-to-vehicle (V2V) network simulations in a two-lane highway scenario using a traffic mobility model (TMM). A microscopic simulation model, the car-following model, is used to describe the movement of vehicles in the forward direction and the vehicles are allowed to change lane when necessary. Realistic parameters are used for the simulations to achieve a traffic flow being as realistic as possible. Based on the simulated traffic the positions of all vehicles at each instant are recorded. The position information is then used to identify vehicles which are in LOS, obstructed-LOS (OLOS) or out-of-range (OoR) from a selected vehicle that is assumed to be a transmitter in the case of VANET simulations. Vehicles at each instant are defined either in one of the LOS, OLOS or OoR states. The intensities of vehicles being in each states are logged which can be used to take into account the impact of OLOS in the VANET simulations. The proposed model is straight-forward to implement, gives realistic results and is based on realistic assumptions for the traffic mobility. Analytical expressions for the packet reception probabilities are used together with the models. The results show the importance of including shadowing by other vehicles for realistic performance assessment.


\begin{thebibliography}{99}

\bibitem{Javier2010}
J.~Gozalvez, M.~Sepulcre, and R.~Bauza, ``Impact of the radio channel modeling
  on the performance of {VANET} communication protocols,''
  \emph{Telecommunication Systems}, pp. 1--19, Dec. 2010.
	
\bibitem{TaimoorAWPL13}
T.~Abbas, J.~Karedal, and F.~Tufvesson, ``{Measurement-based analysis: The
  effect of complementary antennas and diversity on vehicle-to-vehicle
  communication},'' \emph{IEEE Antennas and Wireless Propagation Letters},
  vol.~12, no.~1, pp. 309--312, 2013. [Online]. Available:
  \url{http://lup.lub.lu.se/record/3516482/file/3555826.pdf}

\bibitem{TA2012arXiv}
T.~{Abbas}, F.~{Tufvesson}, and J.~{Karedal}, ``{Measurement based shadow
  fading model for vehicle-to-vehicle network simulations},'' \emph{ArXiv
  e-prints}, Mar. 2012.

\bibitem{Boban11}
M.~Boban, T.~Vinhoza, M.~Ferreira, J.~Barros, and O.~Tonguz, ``Impact of
  vehicles as obstacles in vehicular ad hoc networks,'' \emph{Selected Areas in
  Communications, IEEE Journal on}, vol.~29, no.~1, pp. 15--28, Jan. 2011.

\bibitem{Meireles10}
R.~Meireles, M.~Boban, P.~Steenkiste, O.~Tonguz, and J.~Barros, ``Experimental
  study on the impact of vehicular obstructions in VANETs,'' in \emph{2010 IEEE
  Vehicular Networking Conference (VNC)}, dec. 2010, pp. 338--345.

\bibitem{ns-2}
The network simulator - ns-2. [Online]. Available: \url{http://www.isi.edu/nsnam/ns/}

\bibitem{Varga2008}
A.~Varga and R.~Hornig, ``An overview of the omnet++ simulation environment,''
  in \emph{Proceedings of the 1st international conference on Simulation tools
  and techniques for communications, networks and systems \& workshops}, ser.
  Simutools '08.\hskip 1em plus 0.5em minus 0.4em\relax ICST, Brussels,
  Belgium, Belgium: ICST (Institute for Computer Sciences, Social-Informatics
  and Telecommunications Engineering), 2008, pp. 60:1--60:10. [Online].
  Available: \url{http://dl.acm.org/citation.cfm?id=1416222.1416290}

\bibitem{Henderson2006}
T.~R. Henderson, S.~Roy, S.~Floyd, and G.~F. Riley, ``ns-3 project goals,'' in
  \emph{Proceeding from the 2006 workshop on ns-2: the IP network simulator},
  ser. WNS2 '06.\hskip 1em plus 0.5em minus 0.4em\relax New York, NY, USA: ACM,
  2006. [Online]. Available: \url{http://doi.acm.org/10.1145/1190455.1190468}

\bibitem{Barr2005}
R.~Barr, Z.~J. Haas, and R.~van Renesse, ``Jist: an efficient approach to
  simulation using virtual machines: Research articles,'' \emph{Softw. Pract.
  Exper.}, vol.~35, no.~6, pp. 539--576, May 2005. [Online]. Available:
  \url{http://dx.doi.org/10.1002/spe.v35:6}

\bibitem{Weingartner2009}
E.~Weingartner, H.~vom Lehn, and K.~Wehrle, ``A performance comparison of
  recent network simulators,'' in \emph{Communications, 2009. ICC '09. IEEE
  International Conference on}, 2009, pp. 1--5.

\bibitem{SUMO2011}
M.~Behrisch, L.~Bieker, J.~Erdmann, and D.~Krajzewicz, ``{SUMO} - simulation of
  urban mobility: An overview,'' in \emph{SIMUL 2011, The Third International
  Conference on Advances in System Simulation}, Barcelona, Spain, Oct. 2011,
  pp. 63--68.

\bibitem{Bando1995}
M.~Bando, K.~Hasebe, A.~Nakayama, A.~Shibata, and Y.~Sugiyama, ``Dynamical
  model of traffic congestion and numerical simulation,'' \emph{Phys. Rev. E},
  vol.~51, pp. 1035--1042, Feb 1995. [Online]. Available:
  \url{http://link.aps.org/doi/10.1103/PhysRevE.51.1035}

\bibitem{Tang2005}
T.-Q. Tang, H.-J. Huang, and Z.-Y. Gao, ``Stability of the car-following model
  on two lanes,'' \emph{Phys. Rev. E}, vol.~72, p. 066124, Dec. 2005.

\bibitem{karedal11}
J.~Karedal, N.~Czink, A.~Paier, F.~Tufvesson, and A.~Molisch, ``{Path loss
  modeling for vehicle-to-vehicle communications},'' \emph{IEEE Transactions on
  Vehicular Technology}, vol.~60, no.~1, pp. 323--328, 2011.

\bibitem{Cheng07}
L.~Cheng, B.~Henty, D.~Stancil, F.~Bai, and P.~Mudalige, ``Mobile
  vehicle-to-vehicle narrow-band channel measurement and characterization of
  the 5.9 {GHz} dedicated short range communication ({DSRC}) frequency band,''
  \emph{Selected Areas in Communications, IEEE Journal on}, vol.~25, no.~8, pp.
  1501 --1516, Oct. 2007.

\bibitem{Chowdhury2000}
D.~Chowdhury, L.~Santen, and A.~Schadschneider, ``Statistical physics of
  vehicular traffic and some related systems,'' \emph{Physics Reports}, vol.
  329, no. 4–6, pp. 199 -- 329, 2000. [Online]. Available:
  \url{http://www.sciencedirect.com/science/article/pii/S0370157399001179}

\bibitem{Xuan2012}
B.~C. Yiguang~Xuan, ``Identifying lane change maneuvers with probe vehicle data
  and an observed asymmetry in driver accommodation,'' \emph{ASCE Journal of
  Transportation Engineering}, vol.~8, no. 138, pp. 1051--1061, July 2012.

\bibitem{Tang2007}
T.~Q. Tang, H.~J. Huang, S.~C. Wong, and R.~Jiang, ``Lane changing analysis for
  two-lane traffic flow,'' \emph{Acta Mech Sin}, vol.~23, pp. 49--54, 2007.

\bibitem{Yanlin2001}
Y.~Weng and T.~Wu, ``Car-following model of vehicular traffic,'' in
  \emph{Info-tech and Info-net, 2001. Proceedings. ICII 2001 - Beijing. 2001
  International Conferences on}, vol.~4, pp. 101--106 vol.4.

\bibitem{Kurata2003}
S.~Kurata and T.~Nagatani, ``Spatio-temporal dynamics of jams in two-lane
  traffic flow with a blockage,'' \emph{Physica A: Statistical Mechanics and
  its Applications}, vol. 318, pp. 537 -- 550, 2003. [Online]. Available:
  \url{http://www.sciencedirect.com/science/article/pii/S0378437102013766}

\bibitem{Sheng2009}
J.~Li-sheng, F.~Wen-ping, Z.~Ying-nan, Y.~Shuang-bin, and H.~Hai-jing,
  ``Research on safety lane change model of driver assistant system on
  highway,'' in \emph{Intelligent Vehicles Symposium, 2009 IEEE}, June, pp.
  1051--1056.

\bibitem{Filzek2001}
B.~Filzek and B.~Breuer, ``Distance behavior on motorways with regard to active
  safetyء comparison between adaptive-cruise-control (ACC) and driver,'' in
  \emph{International Technical Conference on Enhanced Safety of Vehicles,
  Amsterdam, Netherlands}, 2001.

\bibitem{Capacity1998}
\emph{Special Report 209: Highway Capacity Manual}, 3rd~ed.\hskip 1em plus
  0.5em minus 0.4em\relax copyright by the Transportation Research Board
  National Research Council, Washington, D.C, 1998.

\bibitem{Proakis2006}
J.~Proakis, \emph{{Digital Communications}}, 5th~ed.\hskip 1em plus 0.5em minus
  0.4em\relax McGraw-Hill Science/Engineering/Math, Aug. 2000.

\bibitem{Katrin2011}
K.~Sj\"{o}berg, E.~Uhlemann, and E.~Str\"{o}m, ``How severe is the hidden terminal
  problem in VANETs when using {CSMA} and {STDMA}?'' in \emph{Vehicular
  Technology Conference (VTC Fall), 2011 IEEE}, Sept. 2011, pp. 1 --5.

\end{thebibliography}
\end{document}